\documentclass[fleqn,usenatbib]{mnras}

\usepackage{newtxtext,newtxmath}
\usepackage{hyperref}
\usepackage{multirow}
\usepackage{tabularx}
\usepackage{xcolor}
\usepackage{caption}
\usepackage{subcaption}
\usepackage{bm}
\usepackage[normalem]{ulem}
\usepackage{threeparttable}

\usepackage[T1]{fontenc}

\DeclareRobustCommand{\VAN}[3]{#2}
\let\VANthebibliography\thebibliography
\def\thebibliography{\DeclareRobustCommand{\VAN}[3]{##3}\VANthebibliography}

\usepackage{graphicx}	
\usepackage{amsmath}	

\newcommand{\tron}{\textsc{TRON}}

\newcommand{\psrn}{J0052$-$7551}

\title[Nulling pulsar discovery with TRON]{Mining the time axis with TRON - III. Discovery of a nulling pulsar towards the SMC with MeerKAT}

\author[Blecher, T et al.]{T.~Blecher$^{1,2}$\thanks{E-mail: tariq.blecher@gmail.com}, 
M.~Geyer$^{3}$,
O.~M.~Smirnov$^{1,2,4,6,7}$\thanks{E-mail: o.smirnov@ru.ac.za}, 
A.~Coelen$^{3,5}$,  
I.~Heywood$^{1,2,6,7}$, 
I.~Pelisoli$^{8}$, 
A.~K.~Hughes$^{6}$, 
\newauthor 
J.~S.~Bright$^{6,7}$, 
V.~Prayag$^{3,5}$, 
T.~Myburgh$^{1}$,  
C.~Tasse$^{1,9}$, 
B.~Hugo$^{1,2}$, 
J.~S.~Kenyon$^{1}$, 
H.~L.~Bester$^{1,2}$, 
\newauthor
S.~J.~Perkins$^{1,2}$, 
J.~Dawson$^{1}$, 
V.~G.~Samboco$^{1}$, 
A.~J.~T.~Ramaila$^{1,2}$,  
B.~Ngcebetsha$^{1,2}$,  
\newauthor 
I.~Sihlangu$^{1,2}$, 
C.~Choza$^{6,7}$, 
V.~S.~Dhillon$^{10,11}$,
N.~Castro Segura$^{8}$, 
J.~McCormac$^{8}$
\\
$^{1}$Centre for Radio Astronomy Techniques and Technologies (RATT), Department of Physics and Electronics, Rhodes University, Makhanda, 6140, South Africa\\
$^{2}$South African Radio Astronomy Observatory, Cape Town, 7925, South Africa\\
$^{3}$High Energy Physics, Cosmology \& Astrophysics Theory (HEPCAT) Group, Department of Mathematics and Applied Mathematics, University of Cape Town, \\Rondebosch 7701, South Africa\\
$^{4}$Institute for Radioastronomy, National Institute of Astrophysics (INAF IRA), Via Gobetti 101, 40129 Bologna, Italy\\
$^{5}$Department of Astronomy, University of Cape Town, Private Bag X3, Rondebosch, 7701, Cape Town, South Africa\\
$^{6}$Astrophysics, Department of Physics, University of Oxford, Keble Road, Oxford, OX1 3RH, UK\\
$^{7}$Breakthrough Listen, Astrophysics, Department of Physics, University of Oxford, Keble Road, Oxford, OX1 3RH, UK\\
$^{8}$Department of Physics, University of Warwick, Gibbet Hill Road, Coventry, CV4 7AL, UK\\
$^{9}$GEPI \& ORN, Observatoire de Paris, Université PSL, CNRS, 5 Place Jules Janssen, 92190 Meudon, France\\
$^{10}$Astrophysics Research Cluster, School of Mathematical and Physical Sciences, University of Sheffield, Sheffield, S3 7RH, UK\\
$^{11}$Instituto de Astrof\'{i}sica de Canarias, E-38205 La Laguna, Tenerife, Spain\\
}

\date{Accepted X. Received Y; in original form Z}

\pubyear{2026}

\begin{document}

\label{firstpage}
\pagerange{\pageref{firstpage}--\pageref{lastpage}}
\maketitle

\begin{abstract}
We report on the discovery of a nulling pulsar, PSR J0052$-$7551, in archival MeerKAT L-band observations of the Small Magellanic Cloud (SMC). The discovery was made using the image-plane transient detection pipeline, TRON, which creates a residual image for each time sample and uses a heuristic pruning strategy to identify promising candidates in the resulting temporal image cube. The TRON pipeline triggered on the approximately minute timescale variability associated with the nulling behaviour of a canonical pulsar. Follow-up MeerKAT UHF-band beamformer observations revealed a pulsar with a spin period of 0.55\,s and a measured period derivative of 6.1(6) $\times 10^{-17}$\,s\,s$^{-1}$. The source is seen in projection toward the periphery of the SMC, and has a dispersion measure value of 29.24(2)\,pc\,cm$^{-3}$, which is on the order of the expected Milky Way contribution along the line of sight. We use a Gaussian Mixture Model to fit the nulling behaviour which yielded a nulling fraction of $\sim 32$\%. We find that the null lengths are consistent with an exponential distribution indicating a stochastic process.  Additionally, power spectra of the UHF visibility data reveal an unexplained quasi-periodic feature at $\sim$5--7\,s that cannot be attributed to the source itself, RFI, or standard instrumental effects. Following on from previous image-plane detections of pulsars by the TRON approach, which relied on either scintillation or eclipsing behaviour, this is the first detection triggered by nulling. This discovery confirms TRON's sensitivity to a wide range of astrophysical phenomena of medium timescale duration.
\end{abstract}

\begin{keywords}
   radio continuum: transients -- stars: pulsars: individual -- 
   methods: data analysis -- techniques: interferometric
\end{keywords}

\section{Introduction} 
The impressive sensitivity and stability of the MeerKAT telescope \citep{meerkat} allows for the use of snapshot imaging to detect transient events on timescales ranging from the telescope sampling times ($\sim2-8$~seconds) to the observation length. This timescale regime, from seconds to minutes, bridges the gap between the sub-second periodicities targeted by traditional pulsar searches and the hour-to-day cadences of slow transient surveys, and has remained relatively unexplored despite its potential for uncovering new source classes.

Fast periodic transients are typically discovered by searching high time resolution data for periodic signals, while simultaneously sampling over a range of trial dispersion measure (DM) values to correct for the unknown frequency dependent delays imparted by the Interstellar Medium (ISM). Interferometers, such as LOFAR and MeerKAT, conduct pulsar searches using specialised beamformer backends (that allow for the coherent addition of antenna streams), and can form tens to hundreds of tied-array beams on the sky for enhanced searching, sky coverage and localisation \citep{Haarlem2013, 2021chen}. Using multi-tied-array-beam capabilities, the TRAPUM pulsar search project running on MeerKAT \citep{Stappers2016}, has discovered 367 new pulsars to date\footnote{https://trapum.org/discoveries/, accessed on 20-07-2026.}, including sources in nearby galaxies \citep{Carli2024,Prayag2024,Prayag2025} and in Milky Way globular clusters (e.g. \citealt{Ridolfi2021, Ridolfi2022, Chen2023}).

Complementary to beamformer methods, image-plane transient searches are now used across the wide-field SKA precursors such as MeerKAT, MWA, LOFAR, and ASKAP for the identification of pulsar candidates. Image-plane searches probe a broad range of timescales, from the correlator integration time to multi-epoch cadences that lie outside the short pulsation window probed by beamformed searches. Because detection rests on a source's spectral shape, variability, or polarisation, rather than on a coherent periodic signal, these searches are largely insensitive to the scattering, eclipses, and orbital acceleration that suppress conventional periodicity searches. Eclipsing spider binaries, for example, have been identified from their on/off variability on orbital timescales in cadenced imaging surveys \citep{petrou-redbacks}, while highly scattered pulsars have been recovered from single continuum images via their strong circular polarisation \citep{wang-j1032, sengar-scattered}.
Programs like ThunderKAT \citep{thk-driessen1, thk-driessen2, andersson-citizen-science, thk-andersson}, ASKAP/VAST \citep{askap-wang}, and the LOFAR Transients Key Project \citep{lofar-stewart} have uncovered new populations of variables and transients through both targeted multi-epoch observations and blind searches.

Radio pulsars are typically grouped into slow (mostly isolated) pulsars, and millisecond pulsars (MSPs, with spin periods $\lesssim 10 $\,ms) often found in binary systems. On the whole, radio pulsars systems can exhibit a wide range of periodic and/or quasi-periodic features. Primarily their spin periods, ranging from a few milliseconds to typically a few seconds; and (if in a binary) their orbital periods lasting from hours to days and months. Additionally, some pulsars show intermittent or nulling behaviour, whereby the radio emission switches between \textit{on} and \textit{off} states periodically or quasi-periodically \citep{Backer1970}. This has been thought to correlate with the characteristic age of the pulsar \citep{Ritchings1976, Wang2007}. However, more recent studies have brought these correlations into question, suggesting that traditional methods failed to distinguish weak emission from genuine nulling and therefore produced biased estimates of the fraction of time a pulsar spends in a nulling state \citep{Kaplan_2018, Anumarlapudi2023}. The most extreme cases of nulling are observed for Rapidly Rotating Radio Transients (or RRATs, \citealt{McLaughlin_2006}). RRATs can exist in their \textit{off} state for hours to months. Imprints by the ISM also cause measurable and time-varying changes in the observed flux densities of radio pulsars, with variations ranging from second or minute timescales (for diffractive scintillation) to days or months (for refractive scintillation).

We have developed the TRON pipeline \citep{parrot, mining1} to automatically generate residual time-cubes from radio observations and search these cubes for peaks in emission. The dynamic imaging approach of TRON, while not directly sensitive to the spin frequencies of pulsars, can trigger on medium timescales associated with orbital, intermittent-emission or ISM-related behaviour. The principal challenge in such searches is filtering out real transients from high signal-to-noise artefacts, which in TRON is performed using heuristic filters such as proximity to bright continuum sources and distance from the phase centre. In \citet{parrot}, dynamic imaging was revealed to be sensitive to the effects of a highly intermittent pulsar, dubbed a PARROT\footnote{pulsar with anomalous refraction recurring on odd timescales} pulsar, which exhibited unique refractive properties. Following the development of TRON, the pipeline has detected several MSPs in well-known globular clusters, based on either their orbital and eclipsing timescales (in the case of spider binary systems), or their diffractive scintillation timescales in the case of the highly scintillating (low-DM) globular clusters \citep{47Tuc, mining1}.

In addition to pulsars, other astrophysical objects potentially detectable by TRON include, but are not limited to, long-period pulsars \citep{caleb2022discovery, nhw-transient, nhw-transient2}, pulsating white dwarfs \citep{Marsh2016, pelisoli2023}, and scintillation phenomena \citep{Wu2022, parrot}.

Here we have employed TRON's fast imaging functionality on archival MeerKAT observations towards the SMC which yielded the detection of a time-variable source located at an (RA, Dec) position of 00h52m43.8s, $-$75d51m04s, with a positional uncertainty dominated by the MeerKAT astrometric error of $\sim 1$~arcsec \citep{MGCLS}.

Follow-up observations with MeerKAT have revealed the source to be a $\sim$0.55~s period radio pulsar which exhibits nulling behaviour on longer timescales. This position is approximately 3 degrees away from the center of the SMC, leaving it outside of the core of the irregular galaxy and towards the edge of its extended periphery 
\citep{nidever2011}. The obtained source DM value of 29.2\, pc\,cm$^{-3}$ lies outside the range of DM values of the 14 known SMC pulsars, which spans 71\,pc\,cm$^{-3}$ \citep{Manchester06} to 292\,pc\,cm$^{-3}$ \citep{Carli2024}. The expected DM contribution of the Milky Way towards the SMC, depending on the electron density model employed, is estimated at 30~pc\,cm$^{-3}$ (YMW2016, \citealt{2017yao}),  32~pc\,cm$^{-3}$ (NE2025, \citealt{Ocker2026}) and 42  pc\,cm$^{-3}$ (NE2001, \citealt{2002cordes}). We therefore refer to PSR J0052$-$7551 as a source towards the SMC, but not associated with the SMC.

In Section~\ref{methods}, we detail the observations and data reduction of the archival L-band data, optical and X-ray observations, and the follow-up UHF image-domain and beamformer observations; Section~\ref{results} describes the obtained characteristics of the newly discovered radio pulsar, including its timing, and an analysis of its nulling properties using recent Gaussian Mixture Modeling techniques. In Section~\ref{sec:discussion} we discuss these findings, including the use of Fourier-domain diagnostics to distinguish nulling from long-period transients, as well as the presence of an additional $\sim$5--7~s quasi-periodic feature revealed in our data and currently without a well-established origin. Finally, the conclusions are presented in Section~\ref{conclusion}.

\section{Observations and data reduction}\label{methods}
\subsection{Discovery observation}
The source was originally detected through a blind search on archival MeerKAT observations using the TRON pipeline. The original observation (CaptureBlock ID: 1564951815) was conducted from 2019-08-04 20:51:59 to 2019-08-05 07:12:00~UTC using MeerKAT L-band (900--1670 MHz) with an 8~s sampling time and 4096 frequency channels. The observation schedule followed a standard repeating scan pattern alternating between calibrators and different target fields. J1939$-$6342 was used as the primary calibrator and J0252$-$7104 was used as the secondary gain calibrator. The science targets consisted of ten neighbouring SMC fields, with each target field being observed ten times with scan durations of 288~s amounting to approximately 50~min of total on-source integration time.
\paragraph*{Data reduction.} 
The data were reduced in a conventional manner using the {\sc oxkat} pipeline up until the calibration of direction independent gains \citep{wsclean, cubical, oxkat, tricolour}. The resulting noise level of the residual continuum map was $9.7\,\mu{\rm Jy}$/beam. 
\paragraph*{Initial detection}
Using the TRON pipeline, we generated sampling time (8~s) residual snapshot images of the field and ran a blind transient search on the resulting time-cube. The result of this was the $\sim 20$-$\sigma$ detection of the pulsar which we initially suspected might be a long period transient, based on the $\sim$65~s quasi-periodic intensity fluctuations visible in the 8~s resolution light curve (Figure~\ref{fig:lightcurve_lband}). We were unable to find previous reference to this object in the ATNF pulsar catalogue\footnote{\url{https://www.atnf.csiro.au/research/pulsar/psrcat}} or the NASA/IPAC Extragalactic Database. 

\paragraph*{Optical follow-up}
To place constraints on the existence of an optical counterpart to the source, we observed the field with ULTRACAM \citep{dhillon2007ultracam} on the night of 2024 December 02 for 20~min.
We used filters $u_s$, $g_s$ and $i_s$, which have similar wavelength coverage to the traditional $ugriz$ filters, but improved efficiency. Our exposure time was 60~s in the $g_s$ and $i_s$ bands, and twice that in the $u_s$ band. The data were reduced using the HiPERCAM pipeline\footnote{\url{https://github.com/HiPERCAM/hipercam}}, and the resulting photometry was calibrated using observations of the standard star GD50 taken on the same night and the {\tt cam$\_$cal} package\footnote{\url{https://github.com/Alex-J-Brown/cam_cal}}. 
We did not find an optical counterpart, and from the obtained distribution of magnitudes, we derived 3-$\sigma$ upper limits to the $u_s$, $g_s$ and $i_s$ magnitudes of 21.7, 23.0, and 22.0, respectively.

\paragraph*{X-ray follow-up}
We obtained a single X-ray follow-up observation with the X-Ray Telescope (XRT) on board the Neil Gehrels \emph{Swift} Observatory \citep{XRTciteAKH}. The 2~ks exposure of J0052$-$7551 (Target ID:~19568) was taken on 2025~Feb~26 (MJD~60732). As the source count rate was unknown, we used photon-counting mode and followed the standard pipeline recommendations when selecting the source and background extraction regions \citep{Evans2007PipelineA,Evans2009PipelineB}. No source was detected, and only one count was found within the source extraction region. We therefore computed a 3-$\sigma$ upper limit on the background-subtracted count rate following \citep{Gehrels1986}. We converted this count-rate limit to an approximate X-ray flux using a simple absorbed power-law model (\texttt{pegpwrlw}) with photon index $\Gamma = 2.0$. Interstellar absorption was modelled using \texttt{tbabs}, adopting the abundances from \citet{wilms2000} and an equivalent hydrogen column density of $N_{\rm H} = 5\times10^{20}\,\mathrm{cm^{-2}}$, obtained from the \emph{Colden: Galactic Neutral Hydrogen Density Calculator}\footnote{\url{https://cxc.harvard.edu/toolkit/colden.jsp}}. Our modelling yields a (3-$\sigma$) upper limit on the 0.5--10~keV X-ray flux of ${\lesssim}\,3\,{\times}\,10^{-13}{\rm\,erg\,s^{-1}\,cm^{-2}}$.

\subsection{Follow-up MeerKAT UHF-band Observations}
We conducted follow-up MeerKAT UHF-band (580--1015 MHz) observations consisting of two 3\,h on-source observations with full polarisation recorded (proposal ID: DDT-20241204-IH-01). We observed with a 2\,s sampling time in interferometric mode and 60\,$\mu$s sampling time in beamformer mode with the PTUSE backend \citep{Bailes2020}. The observations were conducted from 16-Dec-2024/19:12:55 to 16-Dec-2024/22:26:01~UTC (CaptureBlock ID: 1734375186) and from 19-Dec-2024/19:02:27 to 19-Dec-2024/22:15:31~UTC (observation ID: 1734633756). J0408$-$6545 was used as the primary calibrator, J0521$+$1638 was used as the polarisation calibrator, and J0252$-$7104 was used as the secondary gain calibrator. 

\paragraph*{Interferometric data reduction}
The data were reduced using the {\sc oxkat} pipeline \citep{oxkat} through to the second-generation calibration (2GC) stage. The initial reference calibration (1GC) stage applied bandpass, delay, and complex gain corrections derived from observations of the primary and secondary calibrators using {\sc casa} \citep{casa}. The 2GC stage performed one iteration of delay self-calibration using {\sc cubical} \citep{cubical}, with imaging performed using {\sc wsclean} \citep{wsclean} with Briggs weighting (robust $= -0.5$). Radio frequency interference (RFI) was flagged using {\sc casa} and {\sc tricolour} \citep{tricolour}. Each epoch was calibrated independently and the two observations were combined during imaging. The resulting noise level of the residual continuum map for the combined UHF observation is $6.3\, \mu{\rm Jy}$/beam.

\subsection{MeerKAT and Murriyang Pulsar-backend Observations}
MeerKAT beamformer data were captured using the PTUSE-backend during the scans on target, leading to six $\sim$1800~s observations on each of the two observing days. The PTUSE backend records full Stokes information, allowing us to characterise the polarisation of PSR \psrn.  In addition to the original two epochs, three 15~min beamformer-only observations were obtained in August and September 2025 (proposal ID: DDT-20250625-MG-01), enabling us to find an initial timing solution for this pulsar. 

Observations of the PSR~J0052$-$7551 field were also conducted with the Murriyang telescope at the Parkes observatory using the Breakthrough Listen backend \citep{Price2018, Price2021} between December 2024 and April 2025. All Murriyang observations were conducted using the ultra-wide-bandwidth low-frequency receiver (UWL, \citealt{Hobbs2020}), operating from 704 to 4032\,MHz with 100\,$\mu$s time resolution. In addition to the above, we accessed archival search data of the field recorded with the Parkes Multibeam on 2001-06-14 (project P269\footnote {P269: \textit{A deep pulsar survey of the Small Magellanic Cloud},  1997–2001}), using a sample time of 1\,ms. A summary of all the beamformer data products is provided in Table~\ref{tab:bfdata}.

\begin{table}
    \centering
    \begin{tabular}{lllll}
       Date & Duration & Receiver & Frequency & Backend \\
       & (s) & & (MHz)\\
       \hline
       \multicolumn{4}{l}{\textbf{MeerKAT}}\\
       2024-12-16 & 10900 & UHF & 544$-$1088& PTUSE\\
       2024-12-19 & 10898 & UHF & 544$-$1088& PTUSE\\
       2025-08-01 & 900 & UHF & 544$-$1088&PTUSE\\
       2025-08-07 & 900 & UHF & 544$-$1088&PTUSE\\
       2025-09-04 & 900 & UHF & 544$-$1088&PTUSE\\
        \hline
       \multicolumn{4}{l}{\textbf{Murriyang}}\\
       2001-06-14 & 8400& Multibeam &1230$-$1518& FB\_1BIT \\
       2024-12-01 & 4800 &UWL &704$-$4032 & BL$^{*}$ \\
       2025-03-09 & 1800 &UWL &704$-$4032 & BL$^{*}$\\
       
    \end{tabular}
    \caption{MeerKAT and Murriyang beamformer observations of PSR \psrn. $^{*}$Breakthrough Listen.}
    \label{tab:bfdata}
\end{table}

\subsubsection{Pulsar Data Reduction}\label{sec:beamformer_data_reduction}
Initial investigative folds of the first two epochs of UHF PTUSE search mode data, using \texttt{dspsr} \citep{dspsr} with the source coordinates as obtained from the TRON detection in the MeerKAT L-band archival data, and a pulse period of $\sim 65$\,s as estimated from the intensity fluctuations also in the archival data, revealed features at much shorter timescales. Subsequent \texttt{dspsr} folds established these features were associated with the spin period of a $\sim0.5$~s period pulsar exhibiting nulling with timescales up to tens of seconds. Initial improved spin and DM parameters were obtained using \texttt{pdmp} within \textsc{psrchive} \citep{2004Hotan, VanStraten2012}.

Using this folding ephemeris, we created folded pulsar archive files at two time resolutions: 10~s sub-integrations for general analysis, and single-pulse resolution for studying individual pulses. Resulting archive files were further decimated and cleaned using conventional \texttt{pam} routines within \textsc{psrchive}, and \texttt{clfd} \citep{Morello2020}. PTUSE data are polarisation calibrated by the observatory as described in \citealt{Serylak2021}; to this a final frontend correction was applied by running  \texttt{pac -XP} within \textsc{psrchive}.

The MeerKAT folding ephmerides was also applied to the aforementioned historic 2001 Parkes data, allowing us to detect the source in these data.

To obtain a timing solution, a high Signal-to-Noise ratio (S/N) template was created by adding a single 3~h epoch of MeerKAT UHF data (using \texttt{psradd} in \textsc{psrchive}) and then converting this into a noise-free template. We use this template to compute time-of-arrival measurements (ToAs), and its associated uncertainties, using \texttt{pat} in \textsc{psrchive},  for the five UHF PTUSE epochs in Table \ref{tab:bfdata}. Averaging the data for each epoch down to four frequency channels and two sub-integrations (time samples) we obtain 8 ToAs per epoch. A pulsar timing model is then fitted to these ToAs using \textsc{tempo2} \citep{2006hobbs, Edwards2006}, while keeping our best fit source coordinates (with an uncertainty of $\sim 1$~arcsec) as obtained from interferometric imaging. 

We are able to detect the pulsar with $S/N=21$ after averaging the full 2.3~h of historic Parkes Multibeam data from 2001, as well as with $S/N=16-19$ when averaging the full durations of each of the epochs of Murriyang UWL data (see Table~\ref{tab:bfdata}). When computing ToAs from these, the associated high ToA uncertainties, along with the additional model parameters needed to account for timing offsets between observing backends, lead to the ToAs not adding constraints to the timing model. Our best model parameter estimation is from using MeerKAT data independently, and is fully consistent with adding additional Murriyang ToAs.

\section{Results}\label{results}
\begin{figure*}
    \centering
\includegraphics[width=\linewidth]{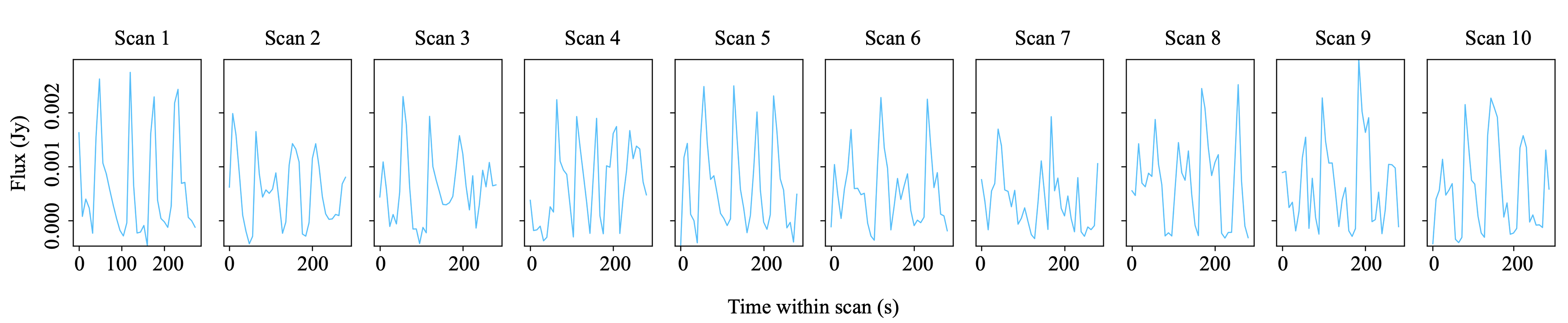}
    \caption{On-source MeerKAT L-band light curve with 8~s sampling time, shown per scan. Each panel displays the flux density as a function of time within the respective scan.}
    \label{fig:lightcurve_lband}
\end{figure*}

In Figure~\ref{fig:lightcurve_lband}, we show the L-band light curve obtained from archival data from which the pulsar was originally detected with a S/N of $17.6$. The spikes in the data correspond to \textit{on} periods, whereas the dips correspond to nulling or \textit{off} episodes.

\begin{figure}
    \centering
    \includegraphics[width=0.8\linewidth]{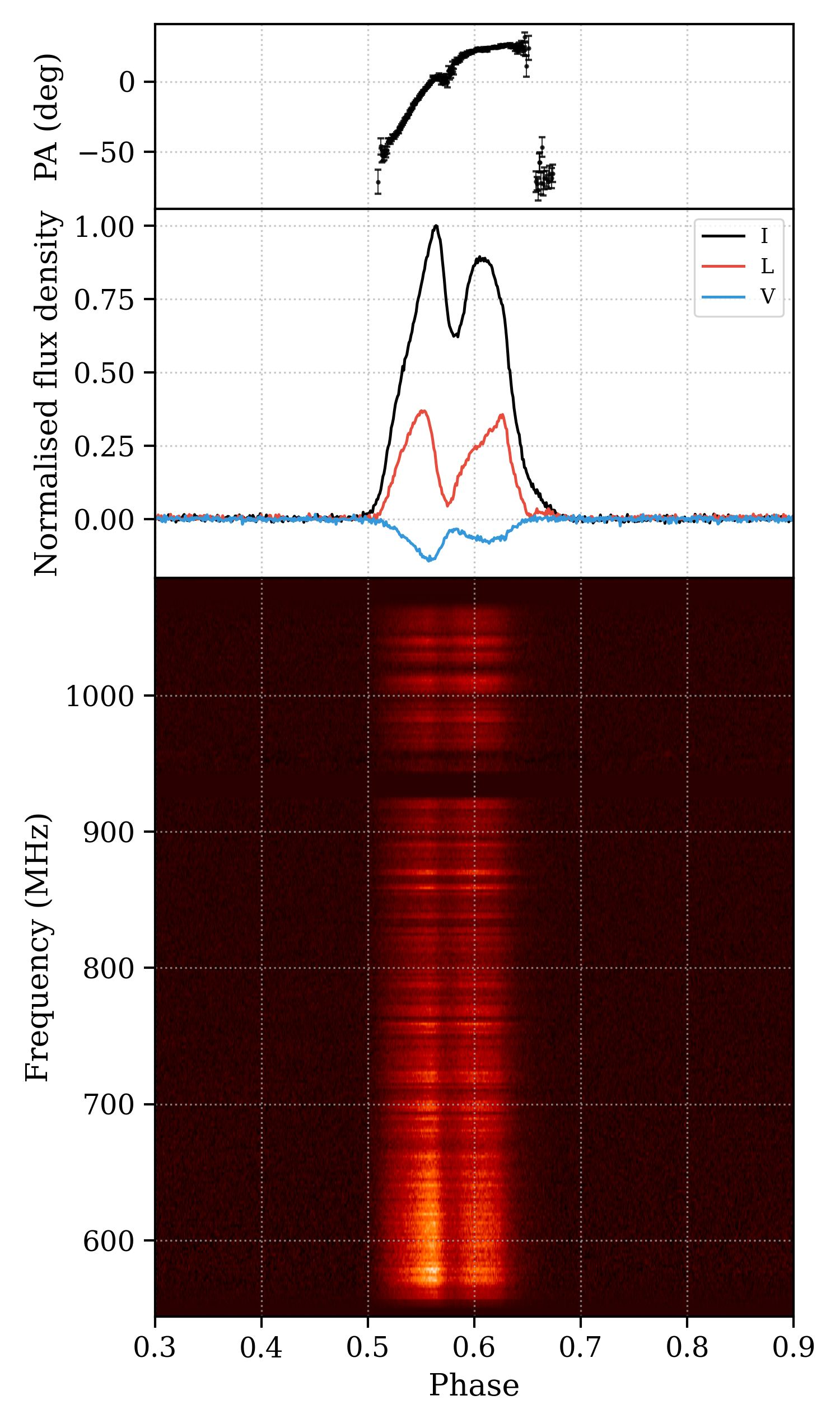}
    \caption{Lower panel: Flux as a function of frequency vs pulse-phase as obtained from the time-averaged beamformer data, across a 3~h UHF observation recorded on Dec 19, 2024 and de-dispersed at DM$=29.2$\,pc\,$\rm{cm}^{-3}$. Middle panel: The averaged pulse profile integrated over frequency. The data were polarisation calibrated according to the procedure described in section~\ref{sec:beamformer_data_reduction} and RM-corrected at RM$=19.94\rm\,{rad}\,{m^{-2}}$ to produce the linear (L, red) and circular (V, blue) polarisation components. Top panel: The corresponding pulsar polarisation angle for the on-pulse region.}
    \label{fig:prof}
\end{figure}

The time-averaged and polarisation-calibrated profile, as obtained using a single 3\,h UHF PTUSE observation, is presented in Figure~\ref{fig:prof}. The data have been corrected for dispersion and Faraday rotation, using a DM of 29.2 pc\,cm$^{-3}$ and a rotation measure (RM) of \mbox{19.94(3) rad\,m$^{-2}$} as estimated from \texttt{rmfit} within \textsc{psrchive} \citep{2004Hotan, VanStraten2012}. This value is consistent with Milky Way foreground models towards the SMC \citep{Mao2008, Livingston2022, Jung2024}, which predict $\mathrm{RM_{MW}} \sim 18$--$31$\,rad\,m$^{-2}$ at the position of \psrn{}. The observed RM therefore does not require a contribution from beyond the Milky Way or from the SMC. Nevertheless, a modest negative RM residual along this sightline cannot be excluded. Such a residual would be consistent with the results of \citet{Livingston2022}, who found that the SMC contribution after foreground subtraction is generally small and predominantly negative, with a median RM of $-10$\,rad\,m$^{-2}$ and a mean coherent line-of-sight magnetic field of $-0.3 \pm 0.1$,$\upmu$G.

We find the average pulse profile has moderate fractional linear polarisation, estimated at 33 per cent, as well as some circular polarisation (Stokes V, 11 per cent). The polarisation position angle (PA), which characterises the projected orientation of the plane of linear polarisation (PA$ = 0.5 \arctan(U/Q)$) exhibits an S-shape curve, characteristic of the basic pulsar rotating vector model \citep{RadhakrishnanCooke1969}. 

\begin{figure*}
    \centering
    \includegraphics[trim=0 30 0 0, clip=True, width=\textwidth]{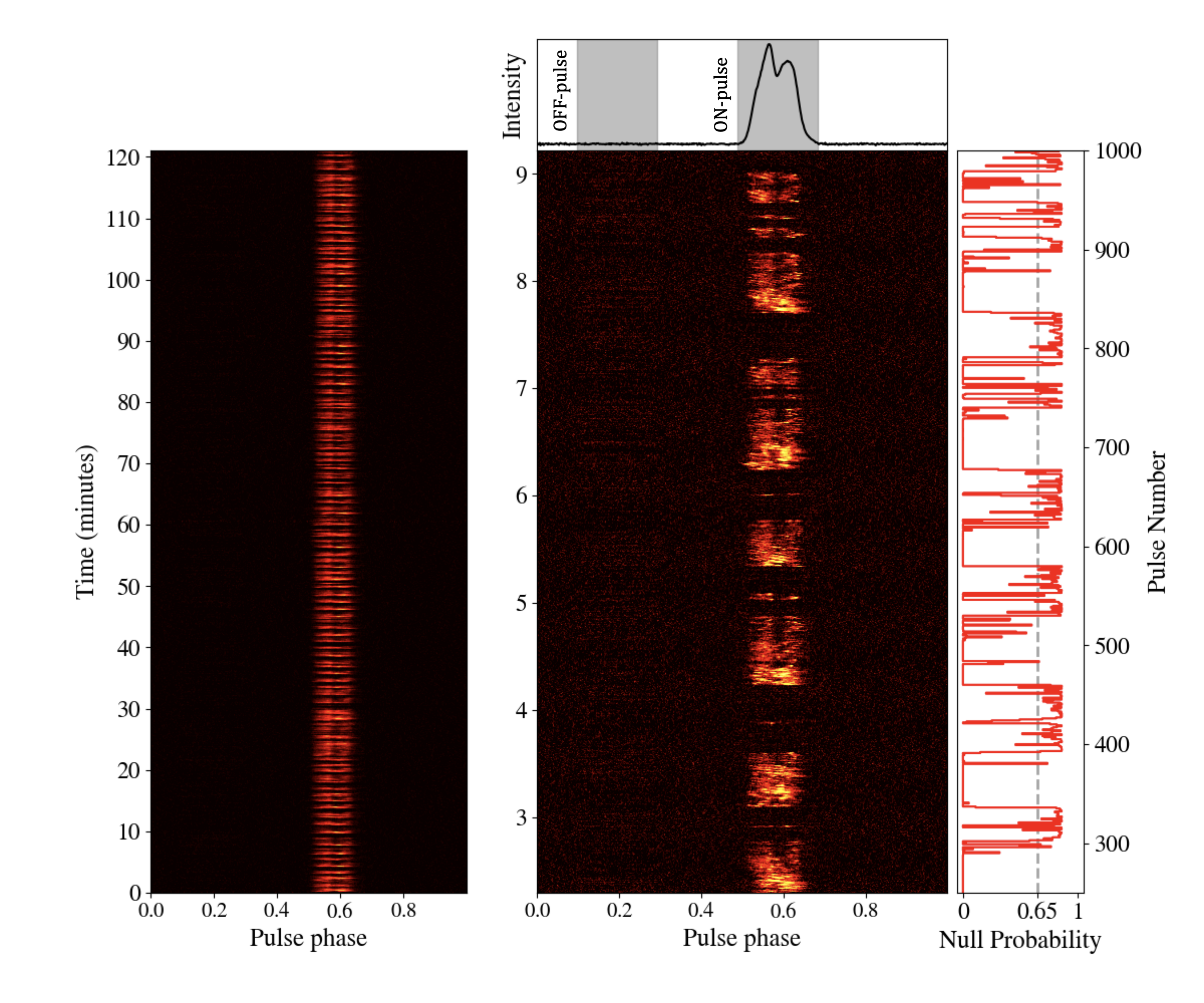}
    \caption{
    Frequency-averaged, DM-corrected Stokes I single-pulse intensities as a function of rotational phase and time for PSR~J0052$-$7551.
    \textbf{Left:} The full 120~min interval used in the analysis of the pulsar's nulling behaviour, highlighting the minute time-scale variability.
    \textbf{Middle:} A zoomed in 7~min section illustrating the variety in length of the consecutive emission and null sequences. The time-averaged pulse profile is shown above this panel, with the chosen OFF-pulse and ON-pulse phase windows indicated by shaded regions.
    \textbf{Right:} Null probabilities computed for each single pulse. The alignment is such that each probability value corresponds directly to the pulse displayed in the same horizontal row of the middle panel, enabling a direct comparison between the observed intensity and its inferred probability of being a null pulse. The grey dashed line indicates the probability threshold used in the classification of null pulses.
    }
    \label{fig:nulling}
\end{figure*}

In Figure~\ref{fig:nulling}, we show examples of frequency-averaged, DM-corrected, Stokes I single-pulse intensities as a function of time and rotational phase. The left-hand side panel shows a 120~min section of the data, where the apparent minute time-scale nulling behaviour is evident. The middle panel provides a zoomed-in view of a shorter $\sim$7~min segment of the same data, highlighting the variability in the length of the consecutive emission and null sequences. Above this zoomed panel, the time-averaged pulse profile is shown, with the selected ON-pulse and OFF-pulse phase windows shaded for reference. The right-hand side panel displays the computed null probabilities for each individual pulse, aligned vertically so that each probability measurement corresponds directly to the pulse shown in the same row of the middle panel.

In addition to the intrinsic nulling behaviour, the pulsar signal is also modulated by interstellar scintillation as it propagates through the ionised ISM. To study the behaviour of the pulsar as a function of both time and frequency across the duration of the observation, we created dynamic spectra with the interferometric data using the {\sc RIMS} package\footnote{\url{https://github.com/saopicc/RIMS}}.

\begin{figure*}
    {\includegraphics[width=0.8\textwidth]{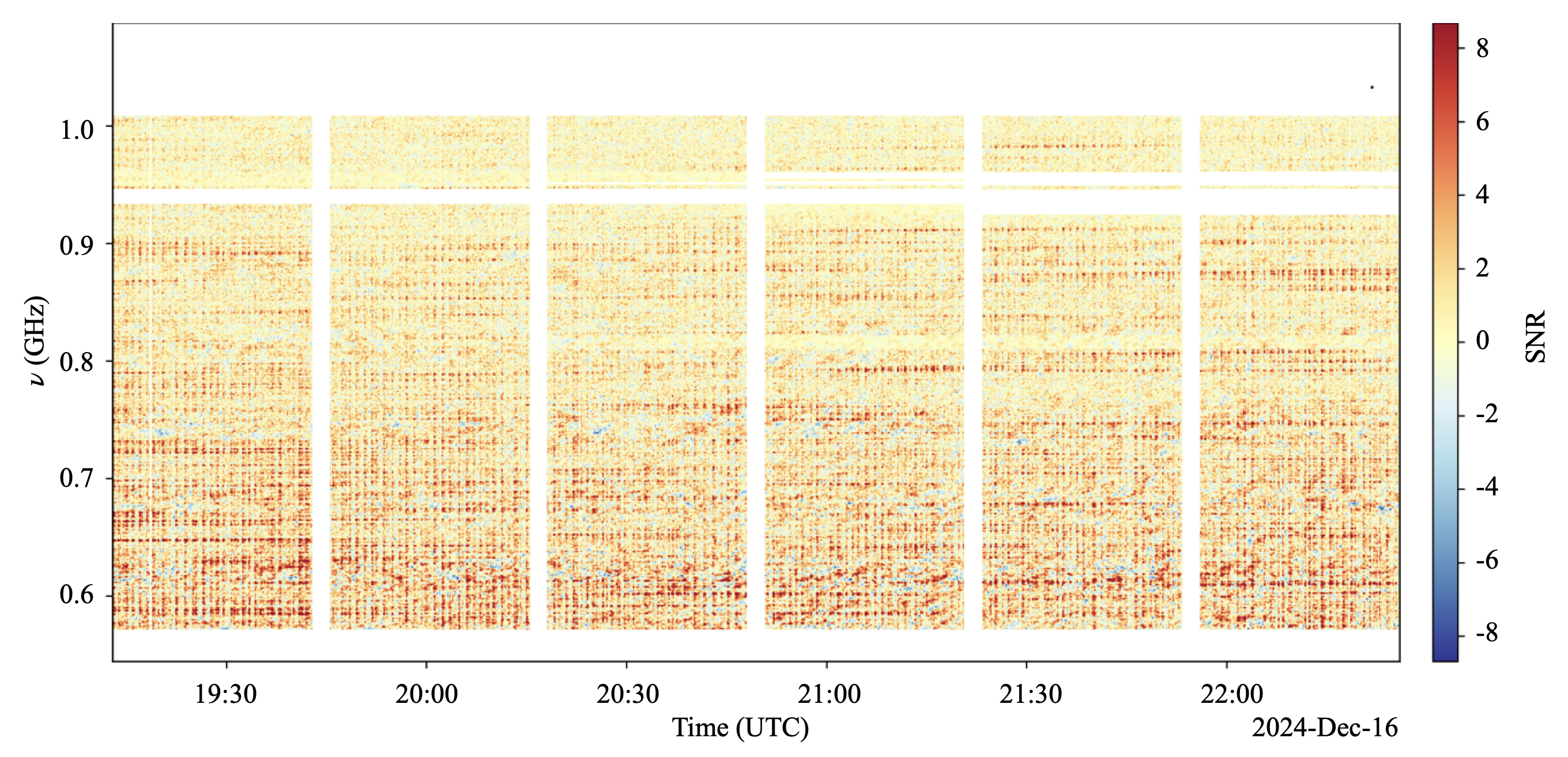}}
    {\includegraphics[width=0.8\textwidth]{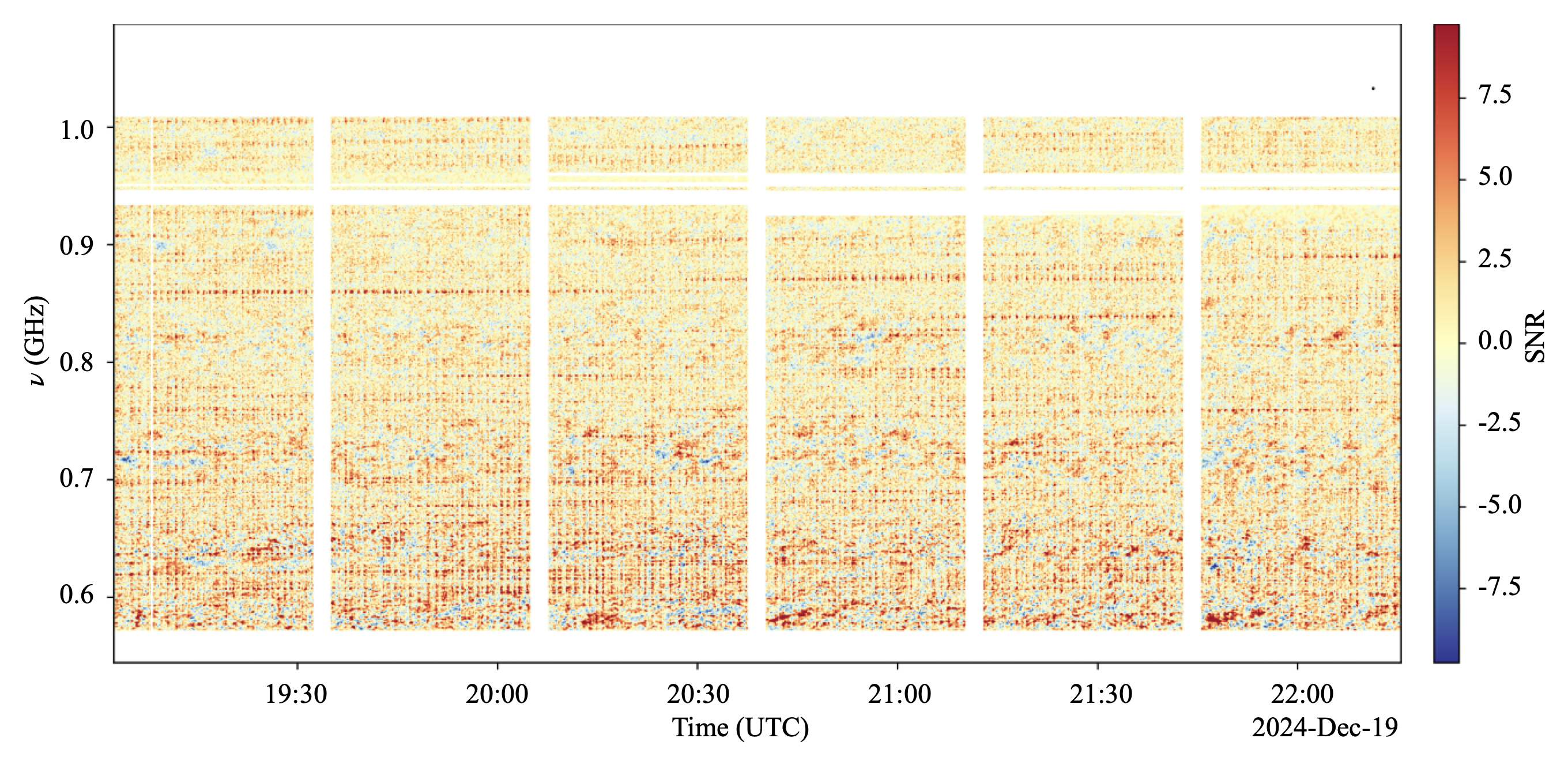}}
    \caption{MeerKAT UHF-band interferometric dynamic spectra, each consisting of 3~h of data. The noise in the dynamic spectra is estimated using the dynamic spectra from off-source parts of the cubes. The dynamic spectra have been smoothed by a Gaussian filter of size (1~MHz, 10~s).
    \label{fig:dynspec_uhf}}
\end{figure*}

The dynamic spectra in Figure~\ref{fig:dynspec_uhf} show interference fringes which are long-lasting (tens of minutes), and that show little change in the peak frequency of each fringe with time. Both of these aspects point to a small effective transverse velocity for the multi-path scattering. The effective velocity is a weighted sum of velocities of source, observer, and scattering material \citep{1998ApJ...507..846C}, and pulsars usually have high values because they typically have high space velocities \citep[mean birth velocity $\sim$400\,km\,s$^{-1}$;][]{Hobbs2005}. This source appears different. That could be because this pulsar happens to have a small value of the relevant space velocity component\footnote{Only one of the three orthogonal space velocity components needs to be small if the scattering is highly anisotropic.}. Alternatively, the pulsar could be at a much greater line-of-sight distance than the scattering material, so that its velocity receives little weight in the sum. 

\begin{figure}
    \centering
    \includegraphics[width=\linewidth]{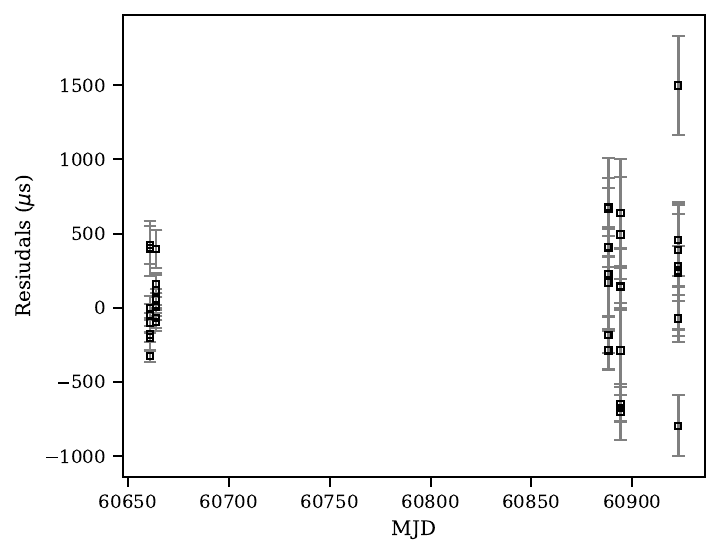}
    \caption{Timing residuals resulting from the best-fit model parameters for  PSR J0052$-$7551 as presented in Table \ref{tab:timing}. The MeerKAT UHF timing baseline spans approximately 8 months. For each observation 8 ToAs were formed, using two sub-integrations and averaging the UHF band to four frequency channels.}
    \label{fig:res}
\end{figure}

A timing analysis, following the procedures described in Section~\ref{sec:beamformer_data_reduction}, allowed us to obtain a phase-connected timing model for this source, the model parameters of which are presented in Table~\ref{tab:timing}. These timing parameters result from keeping the best-fit RA and Dec values fixed at their interferometric positions with uncertainties of $\sim 1$~arcsec in both coordinates, dominated by the MeerKAT astrometric error \citep[e.g.][]{MGCLS}.
The obtained timing residuals in microseconds, after fitting for pulse spin frequency, dispersion measure and the spin frequency derivative, are shown in Figure~\ref{fig:res}. We find a spin period ($P$) of 0.55271150227(6)~s, and a period derivative ($\dot{P}$) of 6.1(6)$\times 10^{-17}$\,s\,s$^{-1}$, providing an estimated characteristic age, $\tau_c$, of $\sim$140 Myr. This estimated age is significantly older than any of the known SMC pulsars for which $\tau_c < 5$\,Myr \citep{Manchester06, Carli2024}.

We note that the Murchison Widefield Array (MWA) SMART survey independently reported the discovery of PSR J$0052-7551$ on 2025 May 8\footnote{mwatelescope.atlassian.net/wiki/spaces/MP/pages/24970773}, providing estimates of the position, DM, and pulse period. The MeerKAT archival detection reported here predates this listing and constrains these parameters to substantially higher precision. With the addition of the follow-up UHF observations, we provide the first measurement of the pulse period derivative.

\begin{table}
\begin{threeparttable}
\centering
\caption{Best fit timing model parameters for PSR J0052$-$7551. Values in parentheses indicate the uncertainty in the last digit.} 
\label{tab:timing}
\begin{tabular}{ll}
\hline\hline
\multicolumn{2}{c}{\textbf{PSR J0052$-$7551}} \\
\hline
MJD range\dotfill & 60660.8 --- 60922.8 \\ 
Data span (yr)\dotfill &  0.72 \\ 
Number of ToAs\dotfill & 40\\
Weighted rms timing residual ($\upmu s$)\dotfill & 251 \\
\hline
\multicolumn{2}{c}{\textbf{Measured quantities}} \\ 
\hline
Right ascension (hh:mm:ss)\dotfill &  00:52:43.8(3)$^{\dagger}$\\ 
Declination (dd:mm:ss)\dotfill & $-$75:51:04(1)$^{\dagger}$\\ 
Pulse frequency, $\nu$ (s$^{-1}$)\dotfill & 1.8092621483(2)\\ 
First derivative of pulse frequency, $\dot{\nu}$ (s$^{-2}$)\dotfill & $-$2.0(2) $\times 10^{-16}$\\
Dispersion Measure (pc cm$^{-3}$)\dotfill & 29.24(2)\\
\hline
\multicolumn{2}{c}{\textbf{Derived quantities}} \\ 
\hline
Pulse period (s)\dotfill& 0.55271150227(6)\\
Pulse period derivative (s s$^{-1}$)\dotfill&6.1(6)$\times10^{-17}$\\
\hline
\end{tabular}
\begin{tablenotes}
\item $\dagger$ As obtained from interferometric imaging.
\end{tablenotes}
\end{threeparttable}
\end{table}

\section{Discussion}\label{sec:discussion}

\subsection{Distinguishing between nulling and long-period sources with interferometric data}\label{sec:distinguish} 
\begin{figure}
    \centering
    \includegraphics[width=\linewidth]{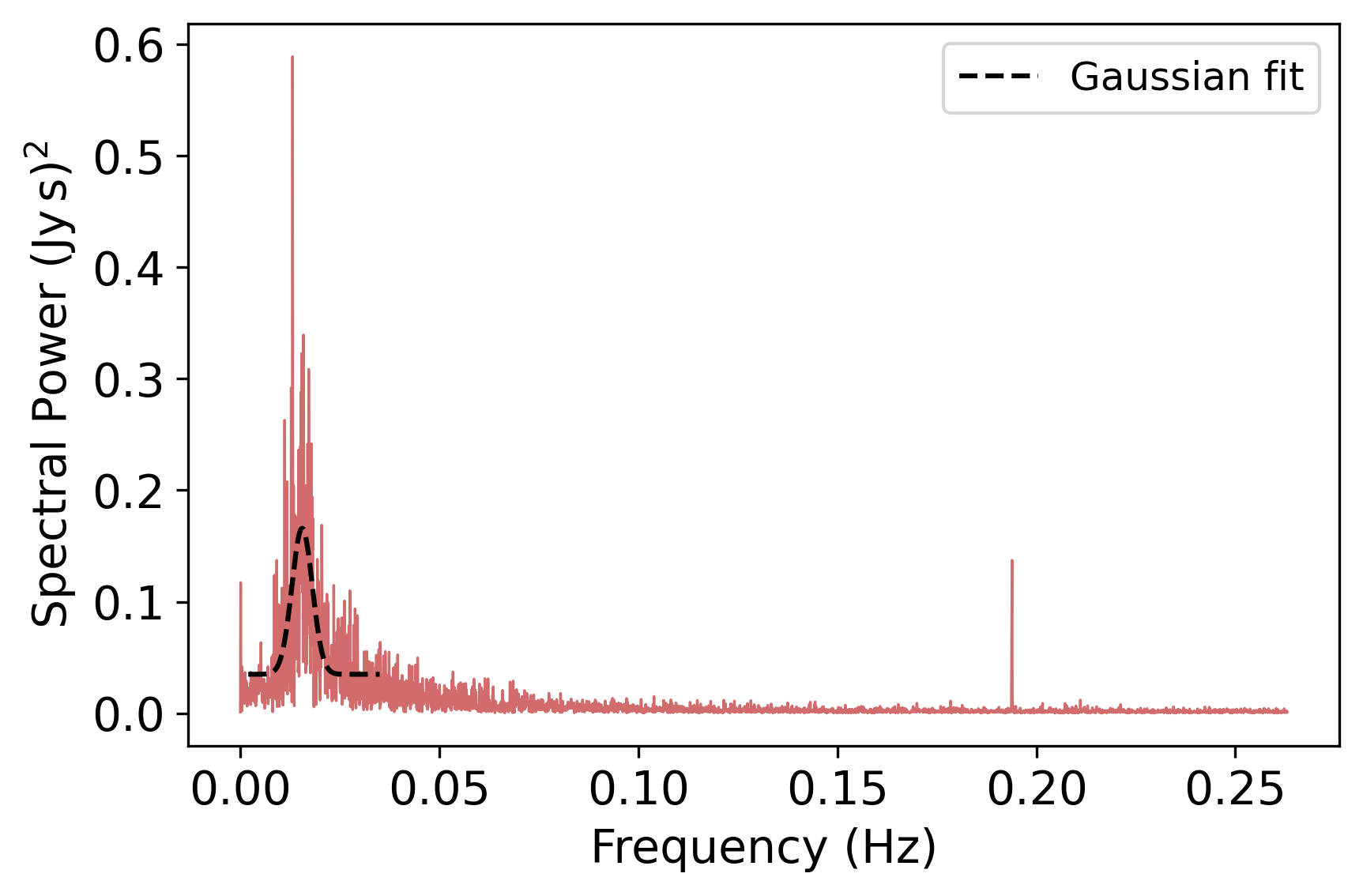}
    \caption{The power spectrum of 3~h of MeerKAT UHF-band interferometric data (2~s integration time). The mean of the original time series is subtracted to remove the zero-frequency power. The broad peak with Gaussian fit corresponds to the nulling behaviour. The sharp peak around 0.19\,Hz shows a $\sim5.2$~s feature discussed in Section~\ref{sec:5s}.
    \label{fig:powerspec_uhf}}
\end{figure}

Using the initial L-band detection observation with a time resolution of 8\,s, the observed minute scale variability was suggestive of a long-period pulsar or transient \citep[e.g.][]{thk-caleb}. This raises the question of whether or not nulling pulsars and long-period sources are distinguishable with the $\sim 8$\,s time resolution available to archival MeerKAT image plane searches. One potential way to distinguish between populations of transients is the width of the power spectrum feature corresponding to the variability, with a more regular transient event having a much larger $Q_{\rm shape}$ factor (where $Q_{\rm shape} = {\rm (mean/FWHM)}$ of a fit Gaussian profile) than a more aperiodic phenomenon such as nulling or moding. Long-period transients, for example, are known to exhibit highly stable periodicities \citep[e.g.][]{thk-caleb}, and would therefore produce a sharp, high-$Q_{\rm shape}$ spectral feature. In Figure~\ref{fig:powerspec_uhf}, we plot the averaged UHF-band power. A Gaussian profile fit to the broad spectral feature yields a centroid frequency of $\sim0.016$\,Hz ($\sim64$\,s) and a full width at half maximum (FWHM) of $\sim0.006$\,Hz, giving $Q_{\rm shape} \approx 2.6$. This low $Q_{\rm shape}$ value shows that the power in the Fourier domain is spread over a wide range of frequencies, reflecting the stochastic nature of the nulling timescales.

\subsection{Nulling}\label{sec:nulling} 
To analyse the nulling properties of the pulsar, we studied single pulse data obtained from the first 120~min of the 3~h observation on 2024 December 16. Increased off-pulse baseline variations, likely linked to the strong observed GSM-band RFI, present in the remaining 60~min of the observation resulted in a contamination of the computed null statistics and were therefore excluded. 

We aim to construct distributions of ON- and OFF-pulse intensities, where the ON-pulse window contains both the pulsar's emission and radiometer noise, while the OFF-pulse window samples only the radiometer noise. The ON- and OFF-pulse phase windows, of equal width, were selected with reference to the averaged pulse profile (see Figure~\ref{fig:nulling}).

The per-pulse off-pulse baseline is subtracted as follows. First, the mean intensity value outside the ON-pulse window was subtracted across the entire pulse phase. Subsequently, a 6th-order polynomial was fitted to the residual baseline, with both the ON- and OFF-pulse windows masked. This polynomial was then subtracted across the full pulse phase to correct the off-pulse baseline variations.

\begin{figure}
    \includegraphics[width=0.48\textwidth, trim=10 10 0 0, clip=True]{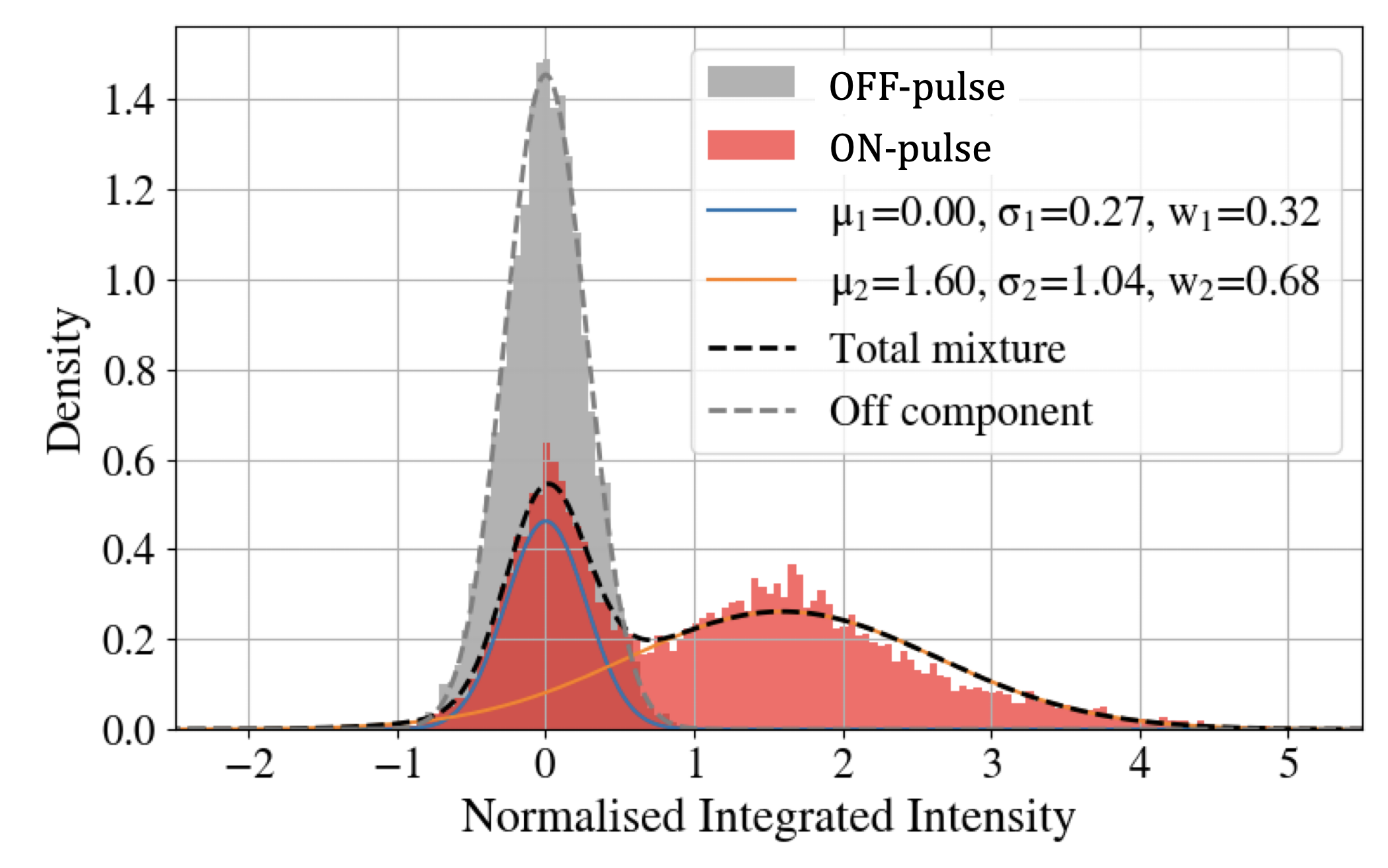}
    \caption{The constructed histograms of the integrated baseline-subtracted single pulse intensities in the ON- and OFF-pulse phase windows, shown in red and grey respectively. The integrated single pulse intensities are normalised with respect to the mean integrated pulse intensity in the ON-pulse phase window. The solid blue (null) and orange (emission) curves indicate the maximum a posteriori individual components from the two-component Gaussian Mixture Model fit, and the dashed black curve their sum, as determined by the MCMC fitting algorithm. }
    \label{fig:ON-OFF_histograms_fit}
\end{figure}

For each pulse, the ON- and OFF-pulse intensities were then computed as the sum of the baseline-corrected flux across their respective windows. Histograms of the recorded ON- and OFF-pulse intensities are presented in Figure \ref{fig:ON-OFF_histograms_fit}, where the intensities have been normalised with respect to the mean integrated pulse intensity in the ON-pulse phase window.

From Figure \ref{fig:ON-OFF_histograms_fit}, we see that the OFF-pulse histogram is well modelled by a single Gaussian component -- as expected of radiometer noise. However, it is evident that the ON-pulse histogram is made up of multiple components, namely the null and emission components. The nulling of the pulsar manifests as an excess of counts with intensities consistent with the OFF-pulse/radiometer noise component.

With the ON- and OFF-pulse histograms constructed, we implement the methods and codes of \citet{Kaplan_2018} to fit a Gaussian Mixture Model (GMM) to the ON- and OFF-pulse histograms. We consider the ON-pulse histogram to consist of a two-component Gaussian mixture parametrised by means $\mu_j$, standard deviations $\sigma_j$, and weights $w_j$, where $j=1$ represents the null component, and $j=2$ the emission component of the ON-pulse histogram respectively. Thus we have that $w_1=\text{NF}$, the nulling fraction of the pulsar. Furthermore, we have the constraint that $\sum w_j=1$, which reduces the number of free parameters in the model to five.

In accordance with the methods of \citet{Kaplan_2018}, we then explore the $\{\mu_j, \sigma_j,w_j\}$ parameter space using a Markov Chain Monte Carlo (MCMC) approach. The best-fit value for the weight of the null component in the ON-pulse histogram, interpreted as the pulsar's nulling fraction, is $\mathrm{NF}=0.315\pm{0.006}$. This nulling fraction is within the broad range of values reported for the known population of nulling pulsars \citep{Wang2007}. The Gaussian fit to the null-pulse component of the ON-pulse histogram is consistent with zero, implying that there is little to no flux in the null state.

A benefit of the GMM methods of \citet{Kaplan_2018} is that the fitted distributions
provide a way to explicitly determine the probability that any individual pulse of a given intensity belongs to either the null or emission component. The probability that a given pulse, 
with intensity $I$, belongs to the null component of the GMM -- the so-called null probability -- is given by, 

\begin{equation*}
    \quad\quad\quad\quad\quad p(j=1|I)=\frac{w_1\mathcal{N}(\mu_1,\,\sigma_1;\, I)}{\sum_{j=1}^2 w_j\mathcal{N}(\mu_j,\,\sigma_j;\, I)},
\end{equation*}
where the $w_j$, $\mu_j$ and $\sigma_j$ are the parameters of the respective Gaussian components, $\mathcal{N}$, in the GMM. The right-hand panel of Figure \ref{fig:nulling} shows the computed null probabilities for a $\sim7$-min section of the analysed data.

By computing the null probabilities for each pulse in the dataset, we are then able to analyse the underlying periodicities in the cycle between the null and emission states of the pulsar. These periodicities were studied by taking the Fourier transform of the computed null probabilities. The resulting power spectrum displayed a broad peak centred at a frequency of $\sim0.015$\,Hz ($\sim66$~s), which is consistent with the results of Figure~\ref{fig:powerspec_uhf}. 

\begin{figure}
    \centering
    \includegraphics[width=0.45\textwidth]{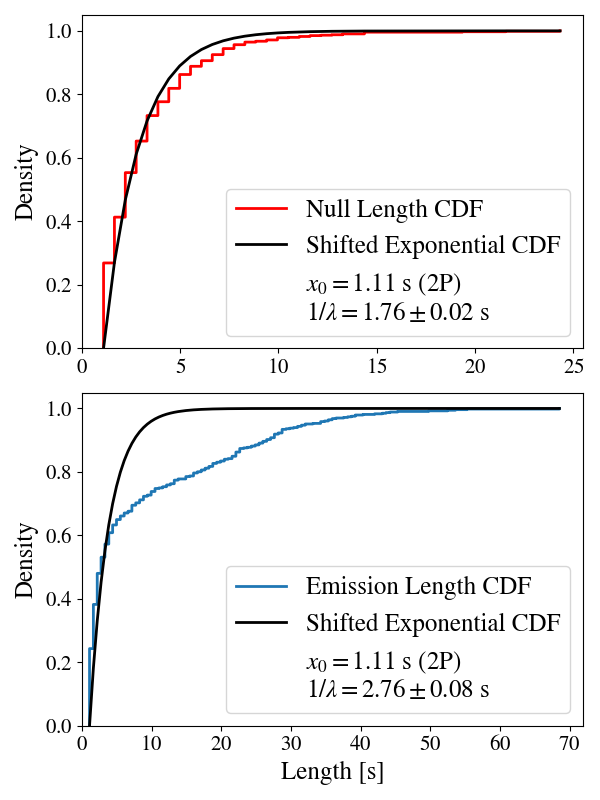}
    \caption{\textbf{Top}: The computed empirical cumulative distribution function (CDF) of null lengths, based on a null probability threshold of 0.65. The corresponding fit of the empirical CDF to the CDF of a shifted exponential distribution is shown with a solid black curve.
    \textbf{Bottom}: The computed empirical CDF of emission lengths, based on the same null probability threshold. The corresponding fit of the empirical CDF to the CDF of a shifted exponential distribution is shown with a solid black curve.}
    \label{fig:null_distribution}
\end{figure}

We also analyse the distribution of null lengths -- the length of time for which the pulsar remains in the nulling state. We define a nulling interval/event to be the number of consecutive pulse periods for which the single pulse intensities have a null probability $>0.65$, with the minimum interval restricted to be of two pulse periods\footnote{The chosen null–probability threshold of 0.65 provides a consistent classification of nulls and yields a nulling fraction of 0.311, in agreement with that derived from the MCMC GMM analysis.}. This minimum length requirement of two pulse periods (2P) introduces a lower bound to the computed null lengths and therefore a shift in the observed distribution.
The empirical cumulative distribution function (CDF) of the computed null lengths is shown in Figure \ref{fig:null_distribution}, together with the best fitting shifted exponential CDF,
\begin{equation*}
    \qquad\qquad\qquad F(x)=1-e^{-\lambda(x-x_0)},\quad x\geq x_0,
\end{equation*}
with $x_0 =2P$. A fit to the shifted exponential CDF yielded a characteristic null length of $1/\lambda = 1.76\pm0.02\,\mathrm{s}$. It should be noted that small deviations between the shifted exponential CDF and the empirical CDF are expected; these arise as a consequence of the intrinsic discretisation imposed by the pulse-by-pulse sampling of the single-pulse data. With these deviations in mind, it can be seen that the empirical CDF closely follows the shifted exponential CDF. This indicates that the mechanism underlying the length of each nulling event is consistent with a Poisson point process, implying that the termination of a null is approximately stochastic in nature.

An analogous analysis was carried out for the emission lengths, defined as consecutive sequences of non-null pulses (again with a minimum length of two pulse periods). In contrast to the null-length distribution, the emission-length CDF, shown in the bottom panel of Figure \ref{fig:null_distribution}, does not closely follow an exponential distribution, indicating that the emission lengths are not well described by a memoryless Poisson point process. This observed deviation from a Poisson process could be connected to the temporal evolution of pulse intensity within each emission state, as seen in Figure \ref{fig:nulling}. In turn, this suggests that the probability of transitioning out of an emission state depends on the length of time spent in the state. 

Such differences in the distributions of null lengths and emission lengths have previously been noted by \citet{Wright}, who likewise interpreted them as evidence that the emission states are likely governed by a process with hysteresis.

\subsection{An unknown $\sim$5--7~s quasi-periodic feature}\label{sec:5s}

\begin{figure}
    \centering
    \includegraphics[width=\linewidth]{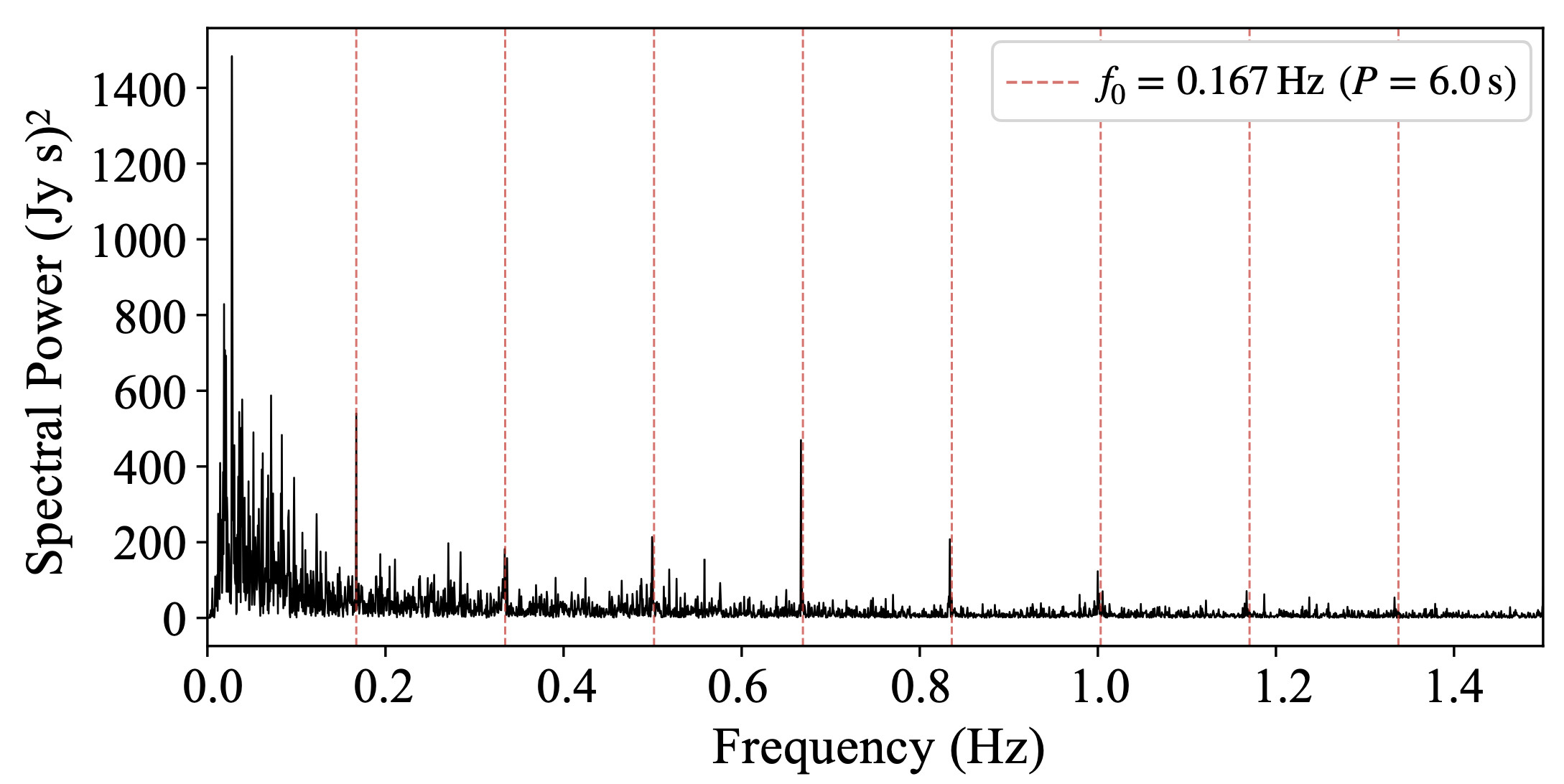}
    \caption{Average power spectrum of the non-dedispersed beamformer data, stacked across two 30~min scans (one per epoch). The higher time resolution beamformer data shows the fundamental peak at $f_0 = 0.167$~Hz ($P \approx 6.0$~s) and its harmonics (dashed red lines) are clearly detected. The strong harmonic content reflects the narrow duty cycle of the repeating broadband features seen in Figure~\ref{fig:beamformer_feature}.}
    \label{fig:beamformer_powerspec}
\end{figure}

\begin{figure*}
    \centering
    \includegraphics[width=\linewidth]{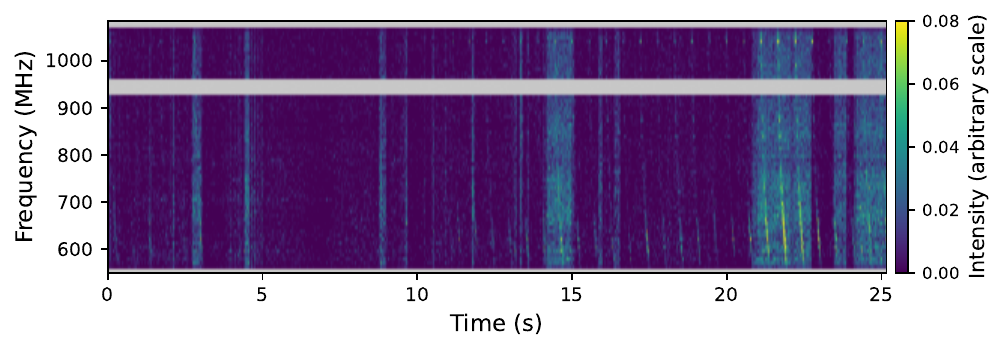}
    \caption{A single continuous $\sim$25~s segment of the non-dedispersed
beamformer dynamic spectrum at the position of PSR \psrn{} (14.8~ms time
resolution, 64 frequency channels spanning the 550--1084~MHz UHF band). The quasi-periodic broadband feature appears as vertical bands recurring every
few seconds (the period varies between $\sim$5 and $\sim$7~s across scans;
in comparison the $\sim$6~s appears as the fundamental in the stacked power spectrum of Figure~\ref{fig:beamformer_powerspec}). The observed features span the full band and are aligned with DM$\sim$0~pc~cm$^{-3}$, inconsistent with the pulsar's DM of
29.2~pc~cm$^{-3}$. Grey blocks mark persistently flagged channels. The flux
scale is arbitrary as beamformer data is typically not flux-calibrated.}
    \label{fig:beamformer_feature}
\end{figure*}

The power spectrum of the full-band light curve of PSR J0052$-$7551 derived from UHF visibilities (Figure~\ref{fig:powerspec_uhf}) shows a distinct narrow peak corresponding to a $\sim$5.2~s mode. To investigate this further, we construct power spectra from each 30~min segment of the beamformer data (with 14.8~ms time resolution). We see a spectral peak at similar periods ($\pm 1$~s) in many but not all of the time segments. The period of the feature varies between $\sim$5 and $\sim$7~s across different 30~min scans. The stacked average power spectrum of two scans (Figure~\ref{fig:beamformer_powerspec}) shows a clear fundamental at 0.167~Hz ($\sim$6.0~s) along with strong harmonics, consistent with the narrow duty cycle of the repeating features. Examining the affected time samples reveals quasi-periodic broadband features (Figure~\ref{fig:beamformer_feature}). Such features in MeerKAT beamformer data are not unusual, and tend to be interpreted as RFI or artefacts and subtracted out, especially using algorithms that implement incoherent beam subtraction from all tied-array beams (Stappers, priv. comm.) However, in our case, having visibility data mystifies the picture: we form dynamic spectra towards other sources in the field, and confirm that this feature only appears in the dynamic spectrum associated with the pulsar's direction. The four plausible origins of the feature are (a) intrinsic to the source, (b) RFI-induced, (c) instrumental response, and (d) propagation effect. However, the data support compelling arguments against the first three, and the fourth does not appear to be likely either.

{\bf Could it be intrinsic to the source?} Investigating the beamformer data as dynamic spectra (frequency-resolved time series), we find that the observed longer period feature, shown in Fig.\ref{fig:beamformer_feature} is consistent with DM$\sim 0$\,pc\,cm$^{-3}$, and does not follow the pulsar's characteristic DM 29.2 pc\,cm$^{-3}$ frequency-sweep across the MeerKAT UHF band. This rules out a direct association with the pulsed 0.55\,s emission of PSR J0052$-$7551. It is also worth noting that we observe this feature during times when the pulsar is in its \textit{off} or nulling state. 

\begin{figure}
    \centering
    \includegraphics[width=\linewidth]{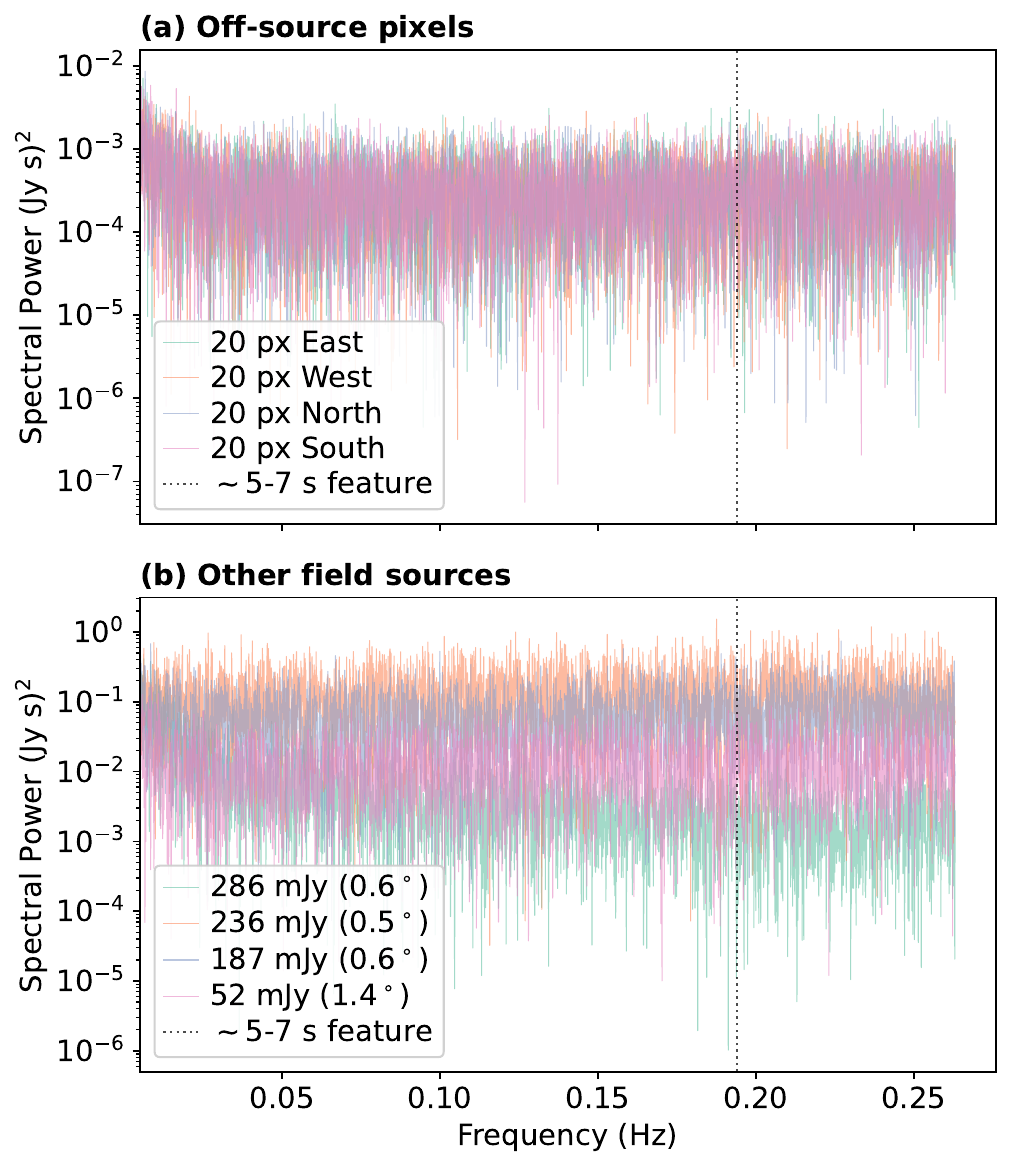}
    \caption{Power spectra of UHF visibility light curves (2~s resolution) for different positions in the field. The dotted line marks the frequency of the $\sim$5--7~s feature detected in the PSR \psrn{} power spectrum (Figure~\ref{fig:powerspec_uhf}). (a) Off-source pixels at offsets of 20~pixels ($\sim$1.6~arcmin) from the pulsar position in each cardinal direction. (b) Four bright field sources at separations of 0.4--0.6~deg. No corresponding feature is present at any comparison position, ruling out additive (RFI) and multiplicative (instrumental) origins.}
    \label{fig:interferometric_powerspec}
\end{figure}

{\bf Could it be RFI?} In snapshot images, RFI manifests itself as one or more fringe-like patterns, depending on the affected antennas and baselines. It is an additive contribution that affects the entire field. Yet power spectra constructed from a number of off-target pixels show no hint of a $\sim$5--7~s feature (Figure~\ref{fig:interferometric_powerspec}a). This implies that any possible additive contribution is localised to the direction of arrival corresponding to the source.

\textbf{Could it be instrumental?} An instrumental response (such as gain) is multiplicative; any uncorrected quasi-periodic feature in the response would imprint itself on the light curves of other sources in the field. Yet power spectra corresponding to other sources in the field (Figure~\ref{fig:interferometric_powerspec}b) show no such feature. This seems to rule out all instrumental effects, except a strongly direction-dependent effect localised to the source direction (phase/pointing centre), but a mechanism for such an effect is very difficult to postulate.

{\bf Could it be a propagation effect?} A propagation effect along the line-of-sight to the source appears to be the last explanation standing, however, neither an atmospheric nor an ISM effect is easy to reconcile with the observation that the effect also persists when the pulsar is in a null state.

It is perfectly possible for such features to have been overlooked in previous observations. Time-domain studies in the visibility domain are a relatively new field; power spectra from 2~s light curves are rarely scrutinised (in fact, ours may well be the first such precedent), and 2~s correlator mode is not commonly used. When seen in beamformer data, such features may easily be dismissed as RFI or artefacts, as noted above. Lacking any plausible mechanism for this observed feature, we have decided to present all available information as is. We would encourage any MeerKAT users who have observed a similar effect to contact the authors.

\section{Conclusion}\label{conclusion}
We have presented the discovery of PSR \psrn, a nulling pulsar found in a blind image-plane search of archival MeerKAT L-band data using the \tron{} pipeline. The source was detected as a variable point source in snapshot residual images towards the SMC, demonstrating the capability of image-plane transient searches to discover pulsars in wide-field interferometric data.
 
Follow-up MeerKAT UHF-band beamformer observations yield a spin period of $\sim 0.55$~s, a period derivative of $6.1 \times 10^{-17}$~s~s$^{-1}$, and a DM of 29.2\,pc\,cm$^{-3}$, consistent with the expected Milky Way contribution along this line of sight. The resulting characteristic age of $\sim 140$\,Myr is significantly older than any of the currently known 14 SMC pulsars. Analysis of the single-pulse data using a Gaussian Mixture Model reveals a nulling fraction of $\mathrm{NF}=0.315\pm{0.006}$, with a characteristic null length of $1/\lambda = 1.76\pm0.02$\,s. The distribution of null lengths follows a shifted exponential distribution consistent with a stochastic Poisson point process, while the distribution of emission lengths is not well described by a Poisson process. Interpreting the null durations as waiting times suggests that the transition out of nulls, to the \textit{on}-state, is a memoryless process, whereas the pulsar's $on$-states, their durations, and the subsequent transition into nulls are determined by the pulsar's more complex underlying internal emission states.

The low $Q_{\rm shape} \approx 2.6$ measured from the Fourier power spectrum of the image-plane light curve reflects the stochastic nature of the nulling timescales, and we propose $Q_{\rm shape}$ as a potential diagnostic for distinguishing between quasi-periodic phenomena such as nulling or moding and the more regular periodicities of long-period transients in image-plane time-series data. Additionally, power spectra of the UHF visibility light curves reveal an unexplained narrow spectral feature at a period of $\sim$5--7~s, whose origin cannot be satisfactorily attributed to intrinsic source emission, RFI, instrumental effects, or propagation phenomena, and which warrants further investigation. This discovery highlights the value of mining archival interferometric observations for transient and variable sources, and underscores the potential of \tron{} as a tool for uncovering new pulsars and transients in the era of MeerKAT and the SKA.

\paragraph*{Data Availability.} The L-band data underlying this article is publicly available via the SARAO archive\footnote{\url{https://archive.sarao.ac.za}}, under proposal ID 1564951815. The UHF-band data will be publicly available from January 2026.

\paragraph*{Acknowledgements.}  The MeerKAT telescope is operated by the South African Radio Observatory, which is a facility of the National Research Foundation, an agency of the Department of Science and Innovation. OMS's, JSK's and VGGS's research is supported by the South African Research Chairs Initiative of the Department of Science, Technology and Innovation and National Research Foundation (grant No. 81737).  MG acknowledges funding from the South African Research Chairs Initiative of the Department of Science, Technology and Innovation and the National Research Foundation of South Africa (grant no. 98626). AC would like to acknowledge support from the South African Radio Astronomical Observatory honour's scholarship programme.

We acknowledge the financial support of the Breakthrough Listen project. Breakthrough Listen is managed by the Breakthrough Initiatives, sponsored by the Breakthrough Prize Foundation. IP acknowledges support from the Royal Society through a University Research Fellowship (URF\textbackslash R1\textbackslash 231496). VSD and ULTRACAM operations are funded by the Science and Technology Facilities Council (grant ST/Z000033/1). We acknowledge the use of the ilifu cloud computing facility – www.ilifu.ac.za, a partnership between the University of Cape Town, the University of the Western Cape, Stellenbosch University, Sol Plaatje University and the Cape Peninsula University of Technology. The ilifu facility is supported by contributions from the Inter-University Institute for Data Intensive Astronomy (IDIA – a partnership between the University of Cape Town, the University of Pretoria and the University of the Western Cape), the Computational Biology division at UCT and the Data Intensive Research Initiative of South Africa (DIRISA). Murriyang, CSIRO's Parkes radio telescope, is part of the Australia Telescope National Facility (https://ror.org/05qajvd42) which is funded by the Australian Government for operation as a National Facility managed by CSIRO. We acknowledge the Wiradjuri people as the Traditional Owners of the Observatory site.
\bibliographystyle{mnras} 
\bibliography{wtf-pulsar} 
 
\label{lastpage}

\end{document}